\documentclass[useAMS,usenatbib]{mn2e}

\def\spose#1{\hbox to 0pt{#1\hss}}
\def\simlt{\mathrel{\spose{\lower 3pt\hbox{$\mathchar"218$}}
     \raise 2.0pt\hbox{$\mathchar"13C$}}}
\def\simgt{\mathrel{\spose{\lower 3pt\hbox{$\mathchar"218$}}
     \raise 2.0pt\hbox{$\mathchar"13E$}}}

\author[Humphrey et al.]{A. Humphrey$^{1,2}$, M. Villar-Mart\'\i n.$^{3}$, R. Fosbury$^{4}$, L. Binette$^{1}$, J. Vernet$^{5}$, \newauthor C. De Breuck$^{5}$, S. di Serego Alighieri$^{6}$ \\
$^{1}$Instituto de Astronom\'\i a, UNAM, Ap. 70-264, 04510 M\'exico, DF, M\'exico (ahumphre@astroscu.unam.mx)\\
$^{2}$Dept. of Physical Sciences, University of Hertfordshire, College Lane, Hatfield, Herts AL10 9AB, UK\\
$^{3}$Instituto de Astrof\'\i sica de Andaluc\'\i a (CSIC), Aptdo. 3004, 18080 Granada, Spain\\
$^{4}$ST-ECF, Karl-Schwarzschild Str. 2, 85748 Garching bei M\"unchen, Germany\\
$^{5}$European Southern Observatory, Karl Schwarschild Str, 2, D-85748 Garching 
bei M\"unchen, Germany\\
$^{6}$INAF-Osservatorio Astrofisico di Arcetri, Largo Enrico Fermi 5, I-50125 Firenze, Italy}
\title[]{Giant Ly$\alpha$ nebulae around z$>$2 radio galaxies: evidence for infall}

\begin{document}

\pagerange{\pageref{firstpage}--\pageref{lastpage}} \pubyear{2006}

\maketitle

\label{firstpage}

\begin{abstract}
We present an investigation into the possible relationship between side-to-side asymmetries of powerful radio galaxies at high redshift, with the goal of understanding the geometry, orientation and gas dynamics of these sources.  Our sample consists of 11 radio galaxies at 2.3$\le$z$\le$3.6 previously known to have giant, kinematically quiescent nebulae.  We identify several correlated asymmetries: on the side of the brightest radio jet and hotspot (i) the redshift of the kinematically quiescent nebula is highest, (ii) Ly$\alpha$ is brighter relative to the other lines and continuum, (iii) the radio spectrum is flattest and (iv) the radio structure has its highest polarization.  These asymmetries are not found to be correlated with either the radio arm length asymmetry or the brightness asymmetry of the UV-optical emitting material.  The correlation between the radio brightness asymmetry and the radial velocity of the quiescent gas also appears to be present in powerful radio galaxies with 0$\la$z$\la$1.  

Collectively, these asymmetries are most naturally explained as an effect of orientation, with the quiescent nebulae in infall: this is the first study to distinguish between the rotation, infall, outflow and chaotic motion scenarios for the kinematically quiescent emission line nebulae around powerful active galactic nuclei.  
\end{abstract}


\section{Introduction}

Powerful radio galaxies have prodigious luminosities, allowing the identification of large samples of such objects across vast ranges in redshift (e.g. R\"ottgering et al. 1997).  Their hosts are believed to be massive ellipticals (e.g. McLure et al. 1999) and are biased towards group or cluster environments (Pentericci et al. 2000a).  Powerful radio galaxies are, therefore, uniquely important for cosmological debates.  

The radio structures of powerful radio galaxies are often asymmetric in terms of brightness, polarization, size and spectral index $\alpha$$_{R}$ (I$_{\nu}$ $\propto$ $\nu$$^{\alpha_{R}}$).  Several correlations between these asymmetries are known.  In powerful radio sources with {\it one-sided jets}, the radio lobe containing the jet depolarizes less rapidly with decreasing frequency (Laing-Garrington effect: Laing 1988; Garrington et al. 1988) and has a flatter spectrum (Garrington, Conway \& Leahy 1991).  In addition, Liu \& Pooley (1991a,b) found that the more depolarized lobe has a steeper spectrum.  Moreover, in powerful radio galaxies with {\it weak jets} the shorter lobe depolarizes more rapidly with decreasing frequency (Pedelty et al. 1989a,b).  

The Laing-Garrington effect is most naturally explained as an orientation effect: Doppler beaming raises the brightness of the nearer (approaching) radio jet, and the associated radio lobe is viewed through a lower column of the surrounding magneto-ionized, depolarizing medium.  The remaining asymmetries require a combination of environmental and orientation effects (Dennett-Thorpe et al. 1997).  

The optical emission lines of powerful radio galaxies also show a spatially asymmetric distribution (e.g. Baum et al. 1988).  This asymmetry appears to be related to the radio structures of powerful radio galaxies, at least out to z$\sim$1.8, in the sense that the line emission is brightest on the side of the closer of the two radio lobes to the nucleus (in projection), suggesting that the arm-length asymmetries of powerful radio sources are the result of environmental effects (McCarthy, van Breugel \& Kapahi 1991; but see also Gopal-Krishna \& Wiita 2005).  

Ly$\alpha$ imaging of z$\ga$2 radio galaxies (HzRG hereinafter) and quasars has revealed Ly$\alpha$ to be spatially extended and, in many cases, spatially asymmetric (e.g. Pentericci et al. 2001; Gopal-Krishna \& Wiita 1996; Heckman et al. 1991).  Interestingly, in their sample of high-z quasars, Heckman et al. (1991) report that for the 9/10 sources for which a spatial asymmetry in the Ly$\alpha$ was discernable, the extended Ly$\alpha$ has a higher surface brightness on the side of the brightest radio jet and hotspot.  These authors argued that this is the result of an orientation effect: if the Ly$\alpha$ emitting gas is anisotropically distributed, and aligned with the radio axis, then the Ly$\alpha$ emission from the side of the receding radio lobe is viewed through a higher collumn of neutral Hydrogen and, consequently, is subject to stronger HI absorption.  

Similarly, Gopal-Krishna \& Wiita (2000) have suggested that the Ly$\alpha$ brightness asymmetries observed in many HzRG might also be the result of an orientation effect and, furthermore, that such asymmetries might allow the orientation to be deduced in sources for which the jet-sidedness cannot be determined.   However, neither Heckman et al. nor Gopal-Krishna \& Wiita were able to compare the spatial distribution of Ly$\alpha$ against that of other UV lines, and as such it is not clear that an alternative origin for their Ly$\alpha$ asymmetry can be rejected (such as an intrinsically asymmetrical distribution of gas or extinction effects: see Wilman, Johnstone \& Crawford 2000).  Thus, the origin of the spatial asymmetry of Ly$\alpha$ has, until now, remained unclear.  

Our group has been carrying out a programme of optical and infrared spectroscopy of HzRG, with the dual aim of understanding the formation and evolution of massive elliptical galaxies, and understanding the way in which the nuclear and radio jet activity affects these processes.  

In an investigation into the nature of the giant gaseous haloes of 10 HzRG (Villar-Mart\'\i n et al. 2002, 2003, 2006; Humphrey et al. 2006), we found that kinematically perturbed gas (FWHM and velocity shifts $\ge$1000 km s$^{-1}$) is associated with, and confined by, the radio structures.  We concluded that vigourous jet-gas interactions are responsible for the kinematic properties of this perturbed gas, and that this gas is in outflow.  Moreover, we identified low surface brightness haloes of kinematically quiescent gas extending across the entire observed extent of the line emission, with FWHM and velocity shifts of several hundred km s$^{-1}$, i.e. consistent with gravitational motion in or around a massive elliptical galaxy.  While the kinematic properties of this quiescent gas are consistent with gravitational motion, it was not possible to genuinely distinguish between rotation, infall and outfall scenarios for this gas.  

In this paper we explore the relationship between the side-to-side asymmetries of powerful z$>$2 radio galaxies, with the main goal of developing the asymmetries as diagnostics of the orientation, geometry and gas dynamics of these sources.  We assume a flat universe with $H_{0}$=71 km $s^{-1}$ Mpc$^{-1}$, $\Omega_{\Lambda}$=0.73 and $\Omega_{m}$=0.27. 

\begin{table*}
\centering
\caption{The sample.  Columns: (1) Source name; (2) redshift; (3) Radio arm-length ratio Q, defined as the length of the longest lobe divided by that of the shortest lobe; (4) Radio hotspot surface brightness ratio at 4.7 GHz, defined as the surface brightness of the brightest hotspot divided by that of the (brightest) hotspot on the opposite side of the nucleus; (5) Modulus of the difference in $\alpha$$_{R}$ between the two hotspots (from van Ojik et al. 1996, Carilli et al. 1997, Pentericci et al. 1999, 2000b); (6) Modulus of the maximum difference in radial velocity, in km s$^{-1}$, either side of the nucleus; (7) Lines used to determine kinematic properties of the quiescent gas; (8) References for spectra. a=Villar-Mart\'\i n et al. (in preparation). b=Vernet et al. (2001) and Villar-Mart\'\i n et al. (2003). c=Vernet et al. (2001) and Villar-Mart\'\i n et al. (2002). d=Humphrey et al. (in preparation). e=van Ojik et al. (1996). f=Carson et al. (2001). g=Overzier et al. (2001) and Villar-Mart\'\i n et al. (2003).} 
\begin{tabular}{llllllll}
\hline
Source & z & Q & R & $\Delta$$\alpha$$_{R}$ & $\Delta$vel & Line & Ref\\  
(1) & (2) & (3) & (4) & (5) & (6) & (7) & (8) \\
\hline
TXS 0211-122 & 2.34 & 1.8 & 2.9 & 0.2 & 200$\pm$70 & HeII $\lambda$1640 & b,d\\
B3 0731+438 & 2.43 & 1.3 & 3.8 & 0.3 & 770$\pm$150 & HeII $\lambda$1640 & b\\ 
TXS 0828+193 & 2.57 & 1.2 & 2.6 & $\le$0.1 & 500$\pm$50 & HeII $\lambda$1640 & c\\
TXS 0943-242 & 2.92 & - & 4.5 & $\le$0.1 & 400$\pm$100 & HeII $\lambda$1640 & b\\
MRC 1243+036 & 3.57 & 1.4 & 1.8 & 0.7 & 450 & Ly$\alpha$ & e\\
4C-00.54 & 2.36 & 1.0 & 1.8 & 0.2 & 600$\pm$100 & HeII $\lambda$1640, [OIII] $\lambda$5007 & b,d\\
TXS 1558-003 & 2.53 & 1.3 & 1.9 & 0.2 & 550$\pm$100 & HeII $\lambda$1640 & a,b,d\\
4C+10.48 & 2.35 & $\sim$1.4 & 11 & 1.1 & 510$\pm$50 & [OIII] $\lambda$5007 & d\\
4C+48.48 & 2.34 & 5.5 & 26 & 0.6 & 450$\pm$80 & HeII $\lambda$1640, [OIII] $\lambda$5007 & b,f\\
MRC 2104-242 & 2.49 & 2.4 & 11 & 0.4 & 250$\pm$10 & Ly$\alpha$, HeII $\lambda$1640 & a,g,d\\
4C+23.56 & 2.48 & 1.9 & 2.7 & 0.6 & 650$\pm$100 & HeII $\lambda$1640, [OIII] $\lambda$5007 & b,d\\
\hline
\end{tabular}
\end{table*}

\section{Data}

The main sample used for this investigation is comprised of 11 radio galaxies with z$>$2, previously known to have low surface brightness emission line nebulae with quiescent kinematics (FWHM$<$700 km s$^{-1}$ i.e. consistent with gravitational motion in a massive elliptical galaxy; see Villar-Mart\'\i n et al. 2003).  Eight of these sources are originally from the Leiden and 4C ultra-steep spectrum compendia (e.g. R\"ottgering et al. 1995).  Two sources (TXS 0943-242 and MRC 2104-242) are taken from the MRC 1 Jy complete sample of McCarthy et al. (1996), and B3 0731+438 is from the study of steep spectrum B3 radio sources carried out by McCarthy (1991).  Table 1 gives some basic details of the sample and the spectroscopic data.  

For 9 sources, we use rest-frame UV spectra from the Keck II telescope, observed using the Low Resolution Imaging Spectrometer (LRIS; Oke et al. 1995) under $\sim$0.7\arcsec to $\sim$1.1\arcsec seeing.  See Cimatti et al. (1998), Vernet et al. (2001) and Villar-Mart\'\i n et al. (2002, 2003) for further details.  

Six of this sample were observed using the Infrared Spectrograph and Array Camera (ISAAC; Moorwood et al. 1998) at the Very Large Telescope (VLT hereinafter; programmes 063.P-0391, 64.P-0500 and 65.P-0579).  Their redshifts allow the rest-optical lines [OII], [OIII] and H$\alpha$ to be observed in the J, H and K bands, respectively.  The ISAAC data will be presented by Humphrey et al. (in preparation).  For the ISAAC and LRIS observation, the slit was usually 1\arcsec wide and was oriented along the radio axis (R\"ottgering et al. 1994; Carilli et al. 1997); the instrumental profile (IP hereinafter) of 500-780 km s$^{-1}$ (FWHM).  

We also use results from integral field spectroscopy of the rest UV emission for MRC 2104-242 and TXS 1558-003.  These spectra have a wavelength coverage of ~4150-6200\AA~and an IP of ~180 km s$^{-1}$.  See Villar-Mart\'\i n et al. (2006; and in preparation) for further details.  

We make use of results from the high-resolution NTT Ly$\alpha$ spectrum of MRC 1243+036 presented by van Ojik et al. (1996).  After an intercomparison between our Keck II spectrum and the NTT spectrum and images of van Ojik et al., we find that the NTT spectrum is oriented oppositely to how van Ojik et al. stated: we find the component with a blueshifted velocity of 1100 km s$^{-1}$ lies to the NW of the nucleus, and that the redshift of the kinematically quiescent gas increases from NW to SE.  Published long-slit spectra of MRC 2104-242 (Overzier et al. 2001) and 4C+48.48 (Carson et al. 2001) were also used.  

To complement the spectroscopic results, we use results from radio continuum imaging presented by van Ojik et al. (1996), Carilli et al. (1997), Pentericci et al. (1999) and Pentericci et al. (2000b), from the Very Large Array at 4.7 and 8.2 GHz at resolutions of $\sim$0.45\arcsec and $\sim$0.25\arcsec.  

\section[]{Results}

In order to obtain information on the orientation, geometry and gas dynamics of HzRG, we now examine the possible relationship between the velocity shear/shift of the kinematically quiescent gas and various spatial asymmetries.  Table 2 summarizes the main results.  

The most important result of this paper is that for 11/11 sources, the redshift of the kinematically quiescent gas is highest on the side of the nucleus which has the brightest radio hotspot (Table 2, column 2).  [For velocity curves and 2D spectra see Overzier et al. (2001), Villar-Mart\'\i n et al. (2002, 2003) and Humphrey (2004)].  In almost all cases, this quiescent gas shows a simple shear or shift of several hundred km s$^{-1}$ in radial velocity along the slit.  

For 9/11 sources we rule out the possibility that the observed velocity shear/shifts result from spatial structure within the slit, because the source filled the slit and/or because the magnitude of the observed velocity shear/shift is too large.  Unfortunately, we could not rule out seeing effects for TXS 0211-122 or TXS 0943-242.  For 10/11 sources, the kinematic results were obtained from non-resonant lines (HeII $\lambda$1640 and/or [OIII] $\lambda$5007), and thus are not subject to line-transfer effects.  Due to the lack of kinematic information from other lines for MRC 1243+036, we use the velocity curve of Ly$\alpha$: the high-resolution spectrum of this source shows no clear evidence for absorption features (see van Ojik et al. 1996).  

\begin{figure}
\includegraphics{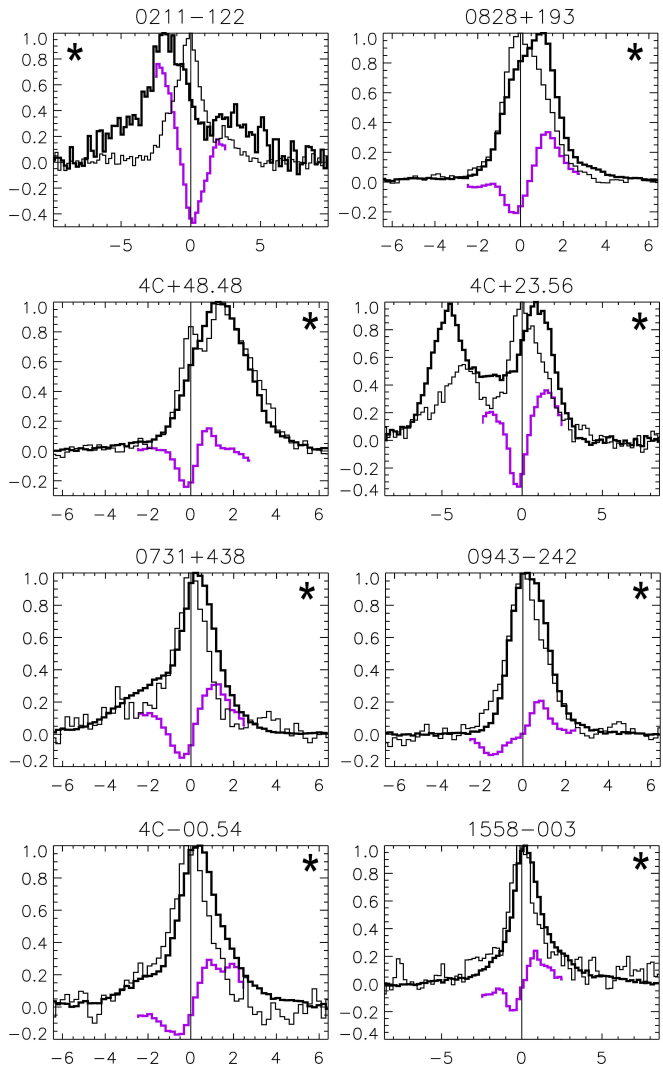}
\vspace{5.65in}\caption{Spatial profile of the Ly$\alpha$ (heavy black line) relative to HeII (TXS 0211-122, TXS 0828+193, 4C+48.48, 4C+23.56) or NV (B3 0731+438, TXS 0943-242, 4C-00.54, TXS 1558-003) (light black line), with the difference between the normalized profiles of Ly$\alpha$ and HeII or NV also shown (heavy violet line).  Horizontal axis: offset along the slit, in arcsec, from the continuum centroid; the peak of the near-IR continuum emission was adopted as the spatial zero.  Vertical axis: normalised flux.  The direction of the brightest radio hotspot is marked by an asterisk (*).  Note the trend for the Ly$\alpha$ emission to be brighter, relative to HeII, on the side of the brightest radio hotspot.}
\end{figure}

We have examined whether this correlation is also present in powerful radio galaxies at lower redshifts (see Appendix A).  For 17/21 of the z$<$2 sources we were able to test, the EELR has its highest redshift on the side of the brightest radio hotspot (Table A1), and this suggests the correlation between the line kinematics and the radio hotspot brightness asymmetry is a common feature of powerful radio galaxies.  

We have compared the spatial distribution of Ly$\alpha$ against that of other UV emission lines and the continuum.  To do this, we extracted from the two-dimensional spectra the spatial profiles of Ly$\alpha$, NV, CIV, HeII and CIII] along the slit.  We summed the flux along the dispersion axis within $\sim$2300 km s$^{-1}$ of the velocity peak, and from this we subtracted the continuum spatial profile extracted from a line-free region.  Figure 1 shows a comparison of the spatial distribution of Ly$\alpha$ against HeII (TXS 0211-122, TXS 0828+193, 4C+48.48 and 4C+23.56) or NV (B3 0731+438, TXS 0943-242, 4C-00.54 and TXS 1558-003); the results do not change significantly if other UV lines or the continuum are used in place of HeII or NV.  

In four sources (TXS 0211-122, TXS 0828+193, 4C+48.48 and 4C+23.56) we find that the spatial peak of the Ly$\alpha$ emission is offset substantially ($\sim$1\arcsec-2\arcsec) from that of the continuum and other emission lines, again towards the brightest radio hotspot (narrow band images of Chambers et al. 1996 and Knopp \& Chambers 1997 confirm this result for 4C+48.48 and 4C+23.56).  In a further four sources (B3 0731+438, TXS 0943-242, 4C-00.54 and TXS 1558-003) we find evidence for a smaller spatial shift ($\sim$0.2\arcsec-0.4\arcsec) from the Ly$\alpha$ peak to the other lines and continuum, again towards the brightest radio hotspot (see table 2 column 3).  Clearly, this spatial shift is not a redenning effect, because it persists when Ly$\alpha$ is compared against lines and continuum at similar wavelengths ($\sim$1240 \AA).  This trend is reminiscent of that indentified by Heckman et al. (1991) in a sample of radio-loud quasars with 2.0$\le$z$\le$2.9, and it appears to be a common feature of powerful, radio-loud active galaxies (Appendix B).

In addition, we indentify several side-to-side radio asymmetries in this sample: in 7/8 cases, the highest linear polarization is measured on the side of the brightest radio hotspot (table 2 column 4), and in 9/9 cases, the brightest hotspot has a shallower radio spectral index than the opposite hotspot (i.e. higher $\alpha$$_{R}$ between 4 and 8 GHz in the observed frame; table 2 column 5).  We note that sources with high hotspot brightness asymmetries have a larger asymmetry in $\alpha$$_{R}$ (table 1).  

We have also examined whether these radio asymmetries show any relationship with the presence of an apparently one-sided radio jet.  For this purpose we define a 'jet' as a linear structure with a length, along the radio axis, of at least two times its width.  We use the 8 GHz images for the reason that they have higher spatial resolution than those at 4 GHz.  Six of our HzRG show a one-sided jet, and for 5/6 of these sources the jet is on the side of the brightest radio hotspot (Table 2, column 6).  Or, to restate this in terms of the other radio asymmetries: the jet is observed to lie on the side of the shallowest $\alpha_{R}$ in 5/5 sources, and on the side of the highest radio polarization in 6/6 sources.  In Appendix C we show that similar correlations between radio asymmetries are also present in the wider sample of HzRG from which our sample was selected.  

It is interesting to note that in our sample, neither the radio arm-length asymmetry Q\footnote{Defined as the length of the longest radio lobe divided by the length of the shortest radio lobe.}, nor the spatial asymmetry in the distribution of non-resonant line emission, are correlated with the asymmetries in radio surface brightness (Table 2; see also McCarthy, van Breugel \& Kapahi 1991), radio polarization, $\alpha$$_{R}$, the Ly$\alpha$ or the velocity structure of the quiescent gas.  However, in sources with high arm-length asymmetries (Q$\ga$1.4) the brightest hotspot tends to be closest to the nucleus, while the opposite is true in the more symmetric sources (Q$\la$1.4).  

\begin{table}
\centering
\caption{Truth table illustrating the relationship between the velocity shear/shift of the quiescent ionized gas, the spatial asymmetry of Ly$\alpha$, and the radio hotspot properties for z$>$2 radio galaxies.  Columns: (1) Source name; (2) Parameter indicating whether the quiescent line emission with the highest redshift is on the side of the nucleus with the brightest radio hotspot; (3) Parameter indicating whether the spatial peak of the Ly$\alpha$ emission is offset, relative to the nucleus, towards the brightest radio hotspot; (4) Parameter indicating whether the highest radio polarization is on the side of the nucleus with the brightest radio hotspot; (5) Parameter indicating whether the brightest radio hotspot has the flattest radio spectrum (highest $\alpha$$_{R}$) between observed frequencies of 4.7 and 8.2 GHz; (6) Parameter indicating, for those HzRG showing a one-sided radio jet, whether the jet is on the side of the brightest radio hotspot; (7) Parameter indicating whether the shortest radio lobe has the brightest hotspot.  A dash (-) denotes that we find no measurable asymmetry in a property or that we lack the data needed to measure an asymmetry.}
\begin{tabular}{lllllll}
\hline
Source & Vel & Ly$\alpha$ & Pol & $\alpha$$_{R}$ & J-R & Q-R \\  
(1) & (2) & (3) & (4) & (5) & (6) & (7) \\ \hline
TXS 0211-122 & 1 & 1 & 1 & 1 & 1 & 1 \\
B3 0731+438 & 1 & 1 & 1 & 1 & 1 & 0 \\
TXS 0828+193 & 1 & 1 & 0 & - & 0 & 0 \\
TXS 0943-242 & 1 & 1 & - & - & - & - \\
MRC 1243+036 & 1 & - & 1 & 1 & 1 & 0 \\
4C-00.54 & 1 & 1 & 1 & 1 & 1 & 0 \\
TXS 1558-003 & 1 & 1 & 1 & 1 & 1 & 0 \\
4C+10.48 & 1 & - & - & 1 & - & 1 \\
4C+48.48 & 1 & 1 & - & 1 & * & 1 \\
MRC 2104-242 & 1 & - & 1 & 1 & - & 1 \\
4C+23.56 & 1 & 1 & 1 & 1 & * & 0 \\
\hline
Total & 11/11 & 8/8 & 7/8 & 9/9 & 5/6 & 4/10 \\
\hline
\end{tabular}
\end{table}

\section[]{Discussion}
In $\S$3 we identified a new set of correlated asymmetries for z$>$2 radio galaxies: on the side of the highest surface brightness radio hotspot, the redshift of the quiescent gaseous nebula is highest, the radio polarization is highest, the radio spectral index $\alpha$$_{R}$ is shallower, the radio jet is brightest, and there is an excess of Ly$\alpha$ flux relative to other UV lines and the continuum.  Clearly, these correlations are unlikely to have occurred by chance.  In this section we aim to discuss the likely physical reasons behind these correlated asymmetries: orientation effects or an environmental asymmetry?  

\subsection[]{Asymmetry in environmental density}
Asymmetries in the environmental density have been invoked by various authors as a way to explain several apparent asymmetries in radio galaxies (e.g. Pedelty et al. 1989b; McCarthy, van Breugel \& Kapahi 1991).  It is, therefore, important to consider whether the correlated asymmetries presented herein might be the result of an asymmetric environment. 

In trying to explain our correlated asymmetries in terms of an environmental asymmetry, we encounter several problems.  Most notably, it is difficult to explain why the velocity of the quiescent gas is correlated with the other asymmetries ($\S$3).  In addition, the apparent correlations with the jet-sidedness, and the lack of any strong correlation with the arm-length asymmetry, reinforce this conclusion.  Therefore, we reject environmental effects in explaining our correlated asymmetries.  

Although we reject environmental effects in the explaining our correlated asymmetries, we do not necessarily reject it in explaining {\it correlated arm-length and line brightness asymmetries} (McCarthy, van Breugel \& Kapahi 1991): indeed there is preliminary evidence for a correlation between the arm-length and non-resonant line brightness asymmetry in our sample (Humphrey et al. in preparation).  We feel it likely that this effect of environmental asymmetry operates {\it alongside} the effect which is responsible for the correlated asymmetries presented herein.  

\subsection[]{Orientation effects}
We now investigate whether our correlated asymmetries can be explained using orientation effects, and we develop a simple scenario involving the viewing angle of the source (see Figure 2).  

\subsubsection[]{One-sided jets}
Relativistic flows can result in large apparent brightness asymmetries (e.g. Ryle \& Longair 1967: Blandford \& K\"onigl 1979) due to Doppler effects.  If the radio axis is oriented at a significant angle to the plane of the sky, and if the jets are moving relativistically as is commonly supposed, then as a result of Doppler-boosting the approaching jet should be brighter than the receding hotspot.  The detection of a one-sided jet, therefore, provides an important test for orientation scenarios.    

As noted in $\S3$ (also Appendix C), we find that on the side of the brightest radio jet, the radio hotspot is also brightest, has a flatter spectrum and has a higher polarization.  This strongly suggests that the asymmetries in polarization, brightness and spectral index are related to the orientation of the radio source.  In $\S$4.2.2-$\S$4.2.5 below, we will discuss in greater detail whether these radio asymmetries are consistent with being the result orientation effects.

\subsubsection[]{Radio hotspot brightness asymmetry R}

If the radio hotspots are also advancing relativistically, and if the radio axis is oriented at a significant angle to the plane of the sky, then the approaching hotspot will be brighter than the receding hotspot as a result of Doppler-boosting.   

We now calculate a simple model in order to test whether Doppler effects can to explain the radio brightness asymmetry of our z$>$2 sample.  For this we use the relation

\begin{equation}R=\left(\frac{1+\beta sin\theta}{1-\beta sin\theta}\right)^{3-\alpha_R}\end{equation}

\noindent where $\beta$ is the hotspot advance speed divided by the speed of light, $\theta$ is the angle of the radio axis relative to the plane of the sky, and $\alpha_{R}$ is the radio spectral index.  We have calculated R for a range in $\theta$ and $\beta$, assuming the radio jets are diametrically opposed and are of equal power.  In the interest of simplicity we have also assumed that the surface brightness of the radio lobe material is negligible in comparison with that of the hotspots (i.e. $\la10$ per cent), and that the hotspots do not have an intrinsic asymmetry in their surface brightness.  

Table 4 shows R for different values of $\theta$ and $\beta$.  (Since we are dealing with radio galaxies, we consider 0$\ge\theta\ge$50$\degr$.)  Our model is indeed able to produce brightness asymmetries of the order of magnitude observed in our sample, but only when $\beta$$\ga$0.1 (compare Table 1 col. 4 against Table 3).  Although the implied hotspot velocities are relatively high, they are broadly consistent with the higher end of hotspot velocity estimates in the literature (e.g. Liu, Pooley \& Riley 1992; Best et al. 1995; Arshakian \& Longair 2000).

\subsubsection[]{$\alpha$$_{R}$ asymmetry}

As we have shown in $\S$4.2.2, Doppler-boosting can have a strong effect on the apparent brightness of the radio hotspots.  The lobe material, on the other hand, is essentially static and thus its brightness should not be affected by relativistic effects; consequently, Doppler-boosting can alter the brightness of the hotspot relative to the radio lobe.  Since the hotspot material emits an intrinsically flatter spectrum than does the lobe material (e.g. Alexander \& Leahy 1987), the greater the relative contribution from the hotspot, the flatter the observed spectrum should be.  Thus, at the position of the approaching hotspot a flatter $\alpha$$_{R}$ should be measured than at the receding hotspot.  

\begin{table}
\centering
\caption{Simple model to test whether Doppler-boosting is able to explain the observed radio brightness asymmetry in our z$>$2 sample.  Columns: (1) Angle of the radio axis to the plane of the sky ($\degr$); (2-6) hotspot brightness asymmetry R calculated for $\beta$=0.05, 0.1, 0.2, 0.3, 0.4.}
\begin{tabular}{llllll}
\hline
$\theta$ & R$_{0.05}$ & R$_{0.1}$ & R$_{0.2}$ & R$_{0.3}$ & R$_{0.4}$ \\  
(1) & (2) & (3) & (4) & (5) & (6)\\ \hline
0  & 1.0 & 1.0 & 1.0 & 1.0 & 1.0\\
10 & 1.1 & 1.2 & 1.4 & 1.6 & 1.9\\
20 & 1.2 & 1.4 & 1.9 & 2.5 & 3.5\\
30 & 1.3 & 1.6 & 2.5 & 3.9 & 6.2\\
40 & 1.3 & 1.8 & 3.2 & 5.8 & 11\\
50 & 1.4 & 2.0 & 4.0 & 8.2 & 17\\
\hline
\end{tabular}
\end{table}

\begin{table}
\centering
\caption{Simple model to test whether Doppler-boosting is able to explain the observed radio spectral index asymmetry in our z$>$2 sample.  Columns: (1) Angle of the radio axis to the plane of the sky ($\degr$); (2-6) radio spectral index asymmetry $\Delta\alpha_{R}$ calculated for $\beta$=0.05, 0.1, 0.2, 0.3, 0.4.}
\begin{tabular}{llllll}
\hline
$\theta$ & $\Delta\alpha_{0.05}$ & $\Delta\alpha_{0.1}$ & $\Delta\alpha_{0.2}$ & $\Delta\alpha_{0.3}$ & $\Delta\alpha_{0.4}$ \\  
(1) & (2) & (3) & (4) & (5) & (6)\\ \hline
0  & 0.00 & 0.00 & 0.00 & 0.00 & 0.00\\
10 & 0.01 & 0.02 & 0.03 & 0.05 & 0.07\\
20 & 0.02 & 0.03 & 0.07 & 0.10 & 0.14\\
30 & 0.02 & 0.04 & 0.09 & 0.14 & 0.20\\
40 & 0.03 & 0.05 & 0.11 & 0.18 & 0.27\\
50 & 0.03 & 0.06 & 0.12 & 0.20 & 0.31\\
\hline
\end{tabular}
\end{table}

In order to examine whether this effect is able to explain quantitatively the observed asymmetry in $\alpha$$_{R}$, we calculate a simple model in which the relative fluxes of the hotspot and lobe varies due to the Doppler-boosting of the hotspot emission.  We assumed: (i) the intrinsic (i.e. un-boosted) surface brightness of each hotspot is a factor of ten higher than that of the lobe material; (ii) the lobe and the hotspot have $\alpha$=-2.5 and $\alpha$=-1.0, respectively; (iii) neither the lobes nor the hotspots have an {\it intrinsic} asymmetry in surface brightness or $\alpha_{R}$.  As in $\S$4.2.2, we consider 0$<\theta<$50$\degr$ and $\beta$=0.05, 0.1, 0.2, 0.3, 0.4.  

In Table 5 we show $\Delta\alpha_{R}$ values for 0$<\theta<$50$\degr$ and $\beta$=0.05, 0.1, 0.2, 0.3, 0.4.  We find that Doppler effects are able to produce $\alpha_{R}$ asymmetries of a similar order of magnitude to those shown by our sample (compare Table 1 col. 5 against Table 4), provided $\beta$$\ga$0.2.  (If we fine-tune the model parameters, and if we also allow for the measurement uncertainties, we can readily reproduce the observed values of $\Delta\alpha_{R}$ for each source.)  Garrington, Conway \& Leahy (1991) came to a similar conclusion for their sample of z$\la$2 radio galaxies and quasars.  

The Doppler-shifting of a curved radio spectrum, and/or a contribution from the Doppler-boosted jet may also contribute to the $\alpha_{R}$ asymmetry, but are unable to explain this asymmetry on their own (see Liu \& Pooley 1991a,b and Dennett-Thorpe et al. 1999, respectively).

\begin{figure*}
\includegraphics{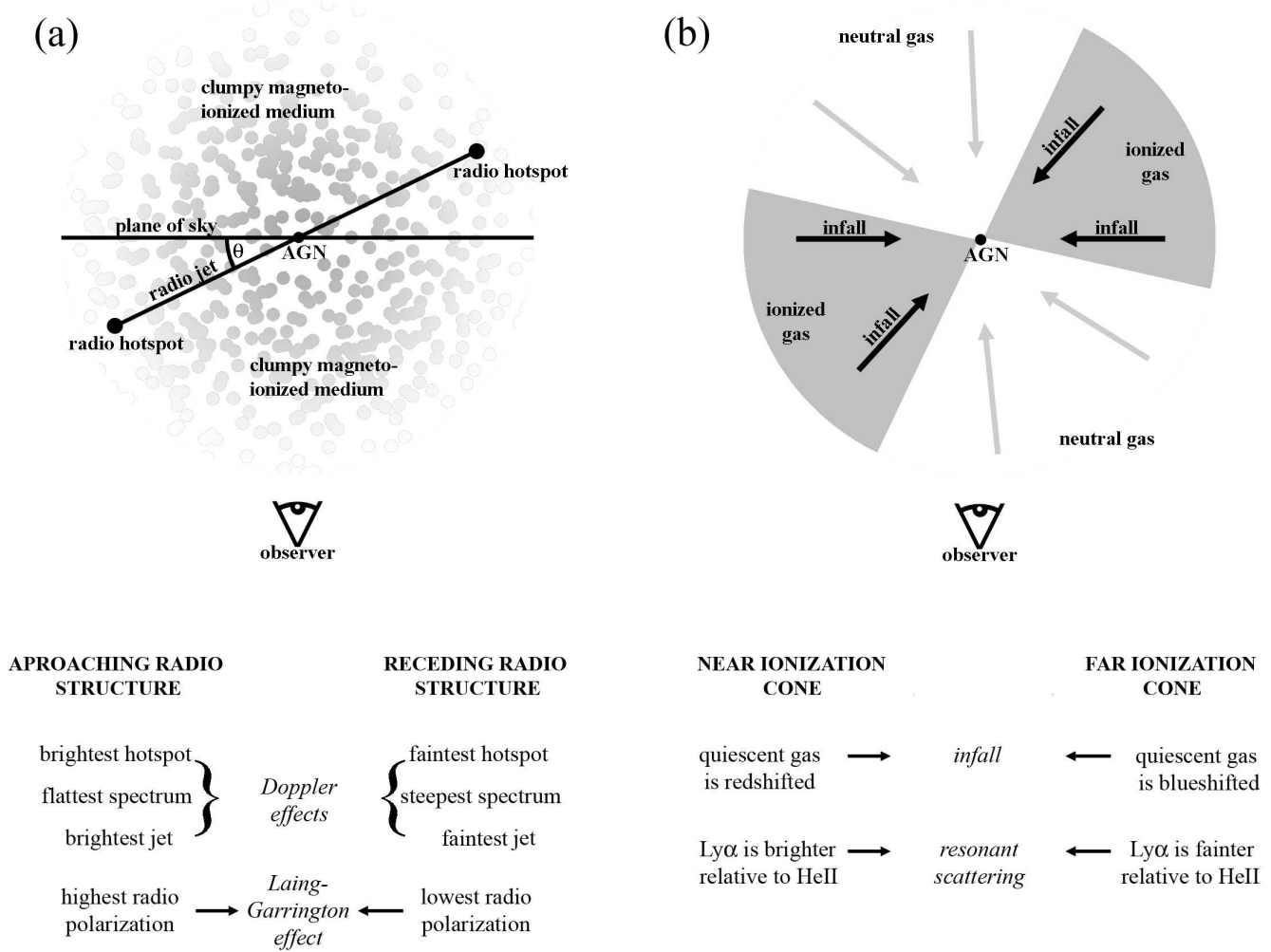}
\vspace{5.65in}\caption{Cartoon to illustrate our orientation scenario.  (a) Radio source asymmetries.   If the radio structure is viewed at an angle, then as a result of Doppler shifting and boosting the approaching radio hotspot is observed to be brighter, and to have a flatter radio spectrum.  In comparison to the receding radio structure, the approaching radio structure is viewed through a lower column of the ambient magneto-ionic depolarizing medium, and consequently is observed to have a higher radio polarization (i.e. is less depolarized); this is analogous to the Laing-Garrington effect.  (b) UV-optical emission line asymmetries.  The ionization bicone shares a common axis with the radio structure, and thus is also viewed at an angle.  The near cone is associated with the approaching radio structure and the far cone is associated with the receding radio structure.   The kinematically quiescent gas is infalling towards the nucleus, and thus the near cone is observed to be redshifted (in projection) relative to the far cone.  The space beyond the ionization bicone contains neutral Hydrogen and, due to the orientation of the bicone/radio axis, the near cone is viewed through a lower column of neutral Hydrogen.  Consequently, the Ly$\alpha$ emission from the near cone suffers less resonant scattering than that from the far cone.  (See text)}
\end{figure*}

\subsubsection[]{Radio polarization asymmetry}

If the radio source is embedded in a magneto-ionized depolarizing medium, and if the radio axis makes a significant angle to the plane of the sky, then the approaching radio lobe would be viewed through a lower column of this depolarizing medium than would be the receding lobe.  Consequently, the approaching lobe would be observed to have a higher degree of polarization (i.e. to be less depolarized) than the receding lobe.  This is analogous to the Laing-Garrington effect (Laing 1988; Garrington et al. 1988), and could explain the polarization asymmetry shown by many of our sample.  

We now test whether this orientation effect is able to explain, quantitatively, the observed polarization asymmetries shown by our sample.  After Burn (1966), we assume that the radio source is viewed through a random distribution of Faraday rotating clouds, and we use the dispersion of the distribution as a measure of Faraday depth.  For a typical polarization asymmetry of a factor of $\sim$4 (Carilli et al. 1997; Pentericci et al. 2000b; Saikia \& Gupta 2003), we then obtain a dispersion of $\sim$1000 cm$^{-3}$ $\mu$G pc.  It is interesting to note that the extreme rotation measures shown by some HzRG (Carilli et al. 1997; Athreya et al. 1998; Pentericci et al. 2000b) also imply Faraday depths of $\sim$1000 cm$^{-3}$ $\mu$G pc.  

This is an order of magnitude higher than typically derived for low-z radio galaxies (e.g. Garrington, Conway \& Leahy 1991), but it does not require an unreasonably large screen size: assuming that the radio source is embedded in a cluster core with a size of $\sim$500 kpc and a gas density of $\sim$0.003 cm$^{-3}$, and assuming a magnetic field strength of $\sim$30 $\mu$G (Pedelty et al. 1989b), the minimum screen size would then be $\sim$10 kpc.  

\subsubsection[]{Correlations between radio asymmetries}
In the preceding sections we have argued that, when considered separately, the asymmetries in brightness, polarization, and spectral index of the radio source can be readily explained in terms of orientation effects.  However, we must also consider whether orientation effects are able to explain the correlations {\it between} the radio asymmetries.

In our orientation scenario, we would expect that on the side of the brightest hotspot: (i) the radio jet is brighter; (ii) the spectral index is flatter; and (iii) the radio polarization is higher (see Figure 2).  These correlations are indeed present in our sample ($\S$3).  We conclude, therefore, that orientation effects provide a natural explanation for our correlations between radio asymmetries.  

If the hotspots advance relativistically (see $\S$4.2.2), then the approaching radio structure should appear longer than the receding structure, due to light travel time effects.  That the arm-length asymmetry Q shows no strong correlation with the other radio asymmetries ($\S$3) is, at first sight, problematic for our orientation scenario (this was also noted by McCarthy, van Breugel \& Kapahi 1991).  

However, this might be explained if environmental effects are the main factor in producing the arm-length asymmetry (as shown by McCarthy, van Breugel \& Kapahi 1991), while the other radio asymmetries are mainly the result of orientation effects.  In support of this, we note that orientation effects are expected to produce a large brightness asymmetry and a small arm-length asymmetry ($\beta$=0.2 and $\theta$=22.5$\degr$ gives R=1.9 and Q=1.1), while an asymmetric distribution of material, on the other hand, is expected to result in a small brightness asymmetry and a large arm-length asymmetry (a density contrast of 5 gives R=1.2 and Q=1.5; see McCarthy, van Breugel \& kapahi 1991).

\subsubsection[]{Ly$\alpha$ asymmetry}
The propagation and escape of resonance line photons can be strongly influenced by the relative geometries of neutral and ionized gas (e.g. Villar-Mart\'\i n, Binette \& Fosbury 1996) and, therefore, the spatial distribution of the resonance line Ly$\alpha$ might be influenced by orientation effects.  We now consider whether the observed spatial asymmetry of Ly$\alpha$ is consistent with orientation effects.  

In the interest of simplicity we assume: (i) that the extended emission line region is photoionized by the active nucleus (e.g. Robinson et al. 1987) and as a consequence has an ionization-bicone morphology (e.g. Jackson, Tadhunter \& Sparks 1998); (ii) that the space outside the bicone contains neutral gas; and (iii) that the ionization-bicone and the radio jets share a common axis, as suggested by the close alignment between the UV-optical line and radio continuum emission (e.g. McCarthy et al. 1987a), and by the observation that the polarization angle of the UV continuum is usually orthogonal to the radio axis (Vernet et al. 2001).  Thus, if the radio axis is at an angle to the plane of the sky, then so too should be the axis of the ionization-bicone.  The near-cone (i.e. associated with the approaching radio lobe) would then be viewed through a lower collumn of H$^{0}$ than the far-cone.  Consequently, the Ly$\alpha$ emission from the near-cone would suffer less resonance scattering, and thus would be brighter, than the Ly$\alpha$ emission from the far-cone (Gopal-Krishna \& Wiita 2005).  We feel that this is the most promising explanation for the observed spatial asymmetry shown by Ly$\alpha$ relative to the other lines (see Figure 2).  Clearly, a detailed quantitative analysis, taking into account transfer effects and also the velocity field of the ambient interstellar medium, would be required in order to evaluate accurately the strength of this effect (see e.g. Hansen \& Oh 2006).  

Within this scenario, we therefore expect that Ly$\alpha$ would be observed to be brighter, relative to the other lines, on the side of the approaching (i.e. brightest) radio hotspot.  The fact that this is true for our sample (Table 2), and moreover for z$\ga$2 radio-loud active galaxies in general (Appendix B; also Heckman et al. 1991), adds further weight in support of our orientation scenario.  

We wish to emphasize that even in the case that these extended emission line regions are shock-ionized, the above orientation scenario should still be valid because shock-ionized gas is expected to be aligned with the radio axis.  

\subsubsection[]{Velocity field of the quiescent gas: evidence for infall}

The dynamics of the kinematically quiescent emission line gas (FWHM$<$700 km s$^{-1}$, i.e. gas unperturbed by jet-gas interactions) has, heretofore, been poorly understood.  Villar-Mart\'\i n et al. (2002, 2003) studied the nature of this gas in detail, but the lack of spatial information in two dimensions prevented these authors from distinguishing between infall, outflow, rotation or, in some cases, chaotic motions.  Even with spatial information in two dimensions, it can be difficult to distinguish unambiguously between these scenarios (Villar-Mart\'\i n et al. 2006).  

As we stated earlier, the most important observational result of this paper is the correlation of the observed velocity field with the brightness, spectral index and polarization of the radio source.  We feel this correlation has by far the greatest diagnostic power of all the correlations discussed herein.  

For our sample, neither rotation nor chaotic motion provides a natural explanation for the observed kinematics of the quiescent gas, since the velocity shifts/shears are correlated with several tracers of geometry, i.e. the spatial asymmetry of Ly$\alpha$, and also the brightness, spectral index and polarization asymmetries of the radio structures -- this correlation clearly should not be present if the quiescent gas is in rotation or in chaotic motion.  

As mentioned above, we assume that the ionization-cones and the radio source share a common axis.  If this axis makes a significant angle to the plane of the sky, and if the gas within the cones is moving radially, then a velocity shear/shift would be observed between the two cones.  In $\S$3 we found that the kinematically emission line quiescent gas has its highest redshift on the side of the brightest radio hotspot: if the brightest hotspot is associated with the approaching radio structure, as we have argued above, then this correlation is most naturally explained by infall of the quiescent gas towards the nucleus (see Figure 2).  

Our sample shows a typical maximum velocity shift of $\sim$400 km s$^{-1}$ (Table 1) at a typical radius of $\sim$40 kpc (Villar-Mart\'\i n et al. 2003).  If we assume our radio galaxies have $\theta\sim$22.5$\degr$, then an infall velocity of $\sim$500 km s$^{-1}$ is implied, which is consistent with models for cosmological infall of gas onto galactic haloes (Barkana 2004).  Adopting a density of $\sim$50 cm$^{-3}$ (e.g. Villar-Mart\'\i n et al. 2003) and volume filling factor of $\sim10^{-5}$ (e.g. McCarthy 1993), we then calculate an order of magnitude rate of gas accretion onto the host galaxy of $\sim$200 M$_{\odot}$ yr$^{-1}$.  

This gas might represent the supply of fuel for the active nucleus, and for the star forming activity seen in some HzRG.  Indeed, the accretion rate calculated above appears to be more than sufficient to fuel both the hidden quasar (which requires $\sim$0.1-10 M$_{\odot}$ yr$^{-1}$: Villar-Mart\'\i n et al. 2003) and the star-formation activity ($\sim$2-60 M$_{\odot}$ yr$^{-1}$: Vernet et al. 2001) of this sample.  

Clearly, this material is not primordial, because it emits metal lines; in fact, there is strong evidence to suggest that these nebulae have around solar metallicity (Vernet et al. 2001; Villar-Mart\'\i n et al. 2003; Humphrey et al., in preparation).  We feel this is consistent with current models for the formation of the hosts of powerful active galaxies (e.g. Granato et al. 2004; Hamman \& Ferland 1999).  Very early in history of the universe, such galaxies are thought to undergo their original rapid mass build-up, accompanied by a high rate of star formation.  This early starburst is likely to be responsible for the synthesis of the metals and for the injection of these metals into the intergalactic medium.  The relatively high level of chemical enrichment observed in the extended haloes, taken together with the presence of a supermassive black-hole and a relatively high dynamical mass (e.g. Villar-Mart\'\i n et al. 2003), suggests that the host galaxy has already experienced this phase of rapid star-formation and mass build-up.  

Considering that the correlation between the radio brightness asymmetry and the radial velocity of the quiescent gas also appears to exist at z$\le$1.1 (Table 3), we suggest that infall may be taking place in almost all powerful radio galaxies at all redshifts.  This is quite remarkable, given the dramatic redshift evolution in the properties of the host galaxy and environment of these sources (see e.g. McCarthy 1993).  It is interesting to note that, at least for the lower-z sources, the host galaxy is already in place (e.g. Mathews, Morgan \& Schmidt 1964).  This adds further weight to the idea that the infall of gas is not necessarily a manifestation of the early evolution of radio galaxy hosts.  

It should be noted that, although we conclude that the {\it quiescent} gas is infalling, this does not contradict the conclusion of Humphrey et al. (2006) that the {\it perturbed} gas is in outflow as a result of strong jet-gas interactions.

\section[]{Summary}
We have conducted an investigation into the side-to-side asymmetries of powerful, high-z radio galaxies, with the main goal of understanding their orientation, geometry and gas dynamics.  We identify several new correlated asymmetries.  Of these, the correlation of the velocity field of the kinematically quiescent halo with radio source parameters is the most surprising, and provides the greatest diagnostic power.  Our main conclusion is that collectively the asymmetries are best explained by orientation effects, with the quiescent nebulae in infall.  We have also carried out a preliminary investigation into side-to-side asymmetries in radio galaxies at lower-z, and draw a similar conclusion.  This is the first study to distinguish between the rotation, infall, outflow and chaotic motion for the quiescent gaseous nebulae gas around powerful active galactic nuclei.  

\section*{Acknowledgments}
AH thanks UNAM for a postdoctoral fellowship.  AH also thanks Sergio Mendoza for stimulating discussions about radio source asymmetries, and thanks Martin Hardcastle for useful discussions about Faraday rotation in radio sources.  We thank the referee, Patrick McCarthy, for useful comments that helped to improve this paper.  The work of MVM was supported by the Spanish Ministerio de Educaci\'on y Ciencia and the Junta de Andaluc\'\i a through the grants AYA2004-02703 and TIC-114.  LB acknowledges support from CONACYT through the grant J-50296.  We would like to thank Laura Pentericci and Chris Carilli for providing us with the radio images of our z$>$2 sample.  We also acknowledge the important contribution Marshall Cohen has made to this project.

\renewcommand\thetable{\bf A\arabic{table}}
\setcounter{table}{0}

\section*{Appendix A: correlation between radio brightness and EELR velocity field at z$<$2}

We have examined, using velocity curves from the literature, whether powerful radio galaxies at z$<$2 also show a correlation between the velocity shift/shear of the quiescent gas and the radio brightness asymmetry.  To avoid sources in which the kinematics may be strongly affected by jet-gas interactions, and to make a germane comparison with our z$>$2 sample, we consider only FRII radio galaxies with radio size $>$200 kpc and EELR FWHM $<$700 km s$^{-1}$.  In addition, we have not considered sources for which the slit was at an angle $\ge$45$\degr$ to the radio axis, or for which two published velocity curves give inconsistent results (i.e. 3C321: compare Robinson et al. 2000 against Tadhunter, Fosbury \& Quinn 1989).  Wherever possible we use radio images with similar rest-frequency and spatial resolution to those used for our z$>$2 sample.  For 17/21 of the z$<$2 sources, we find that the EELR has its highest redshift on the side of the brightest radio hotspot (see Table A1 and references therein).  

\begin{table}
\centering
\caption{Table showing the relationship between the velocity shear/shift of the kinematically quiescent gas and the radio hotspot surface-brightness asymmetry, for powerful radio galaxies at z$<$2.  Columns: (1) Source name; (2) Redshift z; (3) Parameter indicating whether the quiescent line emission with the highest redshift is on the side of the nucleus with the brightest radio hotspot; (4) Observed radio frequency, in GHz; (5) Resolution of the radio observations in kpc; (6) References: a$=$Best, R\"ottgering \& Longair (2000); b$=$Best, Longair \& R\"ottgering (1997); c$=$Inskip et al. (2002); d$=$Best et al. (1999); e$=$McCarthy, Baum \& Spinrad (1996); f$=$Laing (1981); g$=$Riley \& Pooley (1975); j$=$Baum, Heckman \& van Breugel (1990); k$=$Black et al. (1992); l$=$Sol\'orzano-I\~narrea, Tadhunter \& Axon (2001); m$=$Clark et al. (1998); n$=$Villar-Mart\'\i n et al. (1998); o=Tadhunter, Fosbury \& Quinn (1989); p=Reid, Kronberg \& Perley (1999); q=Kronberg, Wielebinski \& Graham (1986); r=Perley, R\"oser \& Meisenheimer (1997).  *$=$although line emission is detected on the E side of the nucleus only, the redshift of this emission increases towards the E, brighter hotspot.}
\begin{tabular}{llllll}
\hline
Source & z & Vel & $\nu$ & Res. & Ref. \\  
(1) & (2) & (3) & (4) & (5) & (6) \\ \hline
3C252 & 1.10 & 1 & 8 & 1.6 & a,b \\
3C356 & 1.08 & 1 & 8 & 1.6 & a,b \\
6C1011+36 & 1.04 & 1 & 8 & 1.6 & c,d \\
3C226 & 0.82 & 1 & 8 & 1.5 & a,b \\
3C265 & 0.81 & 1 & 8 & 1.5 & a,b \\
3C352 & 0.81 & 0 & 8 & 1.5 & l,b \\
3C340 & 0.78 & 1 & 8 & 1.5 & a,b \\
3C441 & 0.71 & 1 & 8 & 1.4 & a,b \\
3C34 & 0.69 & 0 & 8 & 1.4 & l,b \\
3C337 & 0.64 & 1* & 8 & 1.4 & e,b \\
3C330 & 0.55 & 1 & 15 & 3.2 & l,f \\
3C300 & 0.27 & 1 & 5 & 8.2 & e,g \\
3C79 & 0.26 & 0 & 5 & 8.0 & e,g \\
3C171 & 0.24 & 1 & 15 & 2.6 & m,f \\
PKS1932-464 & 0.23 & 1 & 9 & 5.6 & n \\
3C227 & 0.09 & 1 & 8 & 4.1 & j,k \\
PKS0349-27 & 0.07 & 0 & 5 & 2.5 & o,p \\
3C403 & 0.06 & 1 & 8 & 2.7 & j,k \\
3C445 & 0.06 & 1 & 15 & 80 & o,q \\
3C33 & 0.06 & 1 & 5 & 2.3 & j,f \\
PKS0518-45 & 0.03 & 1 & 15 & 5 & o,r \\
\hline
Total & & 17/21 & \\
\hline
\end{tabular}
\end{table}

\renewcommand\thetable{\bf B\arabic{table}}
\setcounter{table}{0}

\section*{Appendix B: Ly$\alpha$ asymmetry and radio brightness asymmetry in the parent sample}

In order to examine whether the correlation between the Ly$\alpha$ and radio asymmetries is a common feature in high-z active galaxies, we have searched the literature for radio galaxies and radio-loud quasars that show a clear asymmetry in the spatial distribution of the Ly$\alpha$ emission.  These sources are listed in Table B1, along with the sources used in the main body of this paper.  Also shown in Table B1 is a comparison of the direction of this asymmetry against that of the other emission lines (where possible), and also against the radio brightness asymmetry.  We find that in most cases, the Ly$\alpha$ emission is brightest on the side of the highest surface brightness radio hotspot.  

\begin{table}
\centering
\caption{Table illustrating the correlation between the spatial asymmetry of the Ly$\alpha$ emission and the radio hotspot brightness asymmetry, for radio galaxies and radio-loud quasars with z$\ga$1.7.  Columns: (1) Source name; (2) Source redshift z; (3) Source classification: radio galaxy (G), broad-line radio galaxy (BG) or radio-loud quasar (Q); (4) Parameter indicating whether Ly$\alpha$ is brightest on the side of the highest surface brightness radio hotspot; (5) For some sources we have been able to compare the distribution of Ly$\alpha$ against that of other lines; this column indicates whether Ly$\alpha$ is brightest {\it relative to other lines} on the side of the brightest radio hotspot; an asterisk (*) denotes that we have not been able to check the Ly$\alpha$ distribution against other lines; (6) References: H91=Heckman et al. (1991); P01=Pentericci et al. (2001); VM06=Villar-Mart\'\i n et al. (2006); vO94=van Ojik et al. (1994); H06=this paper; C90=Chambers et al. (1990); R03=Reuland et al. (2003); GK95=Gopal-Krishna et al. (1995); P97=Pentericci et al. (1997); H07=Humphrey et al. (in preparation); E93=Eales et al. (1993); vO96=van Ojik et al. (1996); MC90=McCarthy et al. (1990); s95=Spinrad et al. (1995); vO97=van Ojik et al. (1997); MC87b=McCarthy et al. (1987b); C96=Chambers et al. (1996); K97=Knopp \& Chambers (1997); M02=Maxfield et al. (2002); PT05=P\'erez-Torres \& De Breuck (2005); I03=Iwamuro et al. (2003).}
\begin{tabular}{llllll}
\hline
Source & z & Class & Ly$\alpha$-R & Ly$\alpha'$-R & Ref\\  
(1) & (2) & (3) & (4) & (5) & (6)\\ \hline
Q 0109+176 & 2.16 & Q & 0 & 0* & H91\\
MRC 0156-252 & 2.09 & G/Q & 1 & 1 & P01; I03\\
TXS 0211-122 & 2.34 & G & 1 & 1 & vO94; H06\\
MRC 0406-244 & 2.44 & G & 0 & - & P01; I03\\
Q 0445+097 & 2.11 & Q & 1 & 1* & H91\\
6C+60.07 & 3.79 & G & 0 & - & R03\\
4C+41.17 & 3.79 & G & 1 & - & C90; R03\\
Q 0730+257 & 2.69 & Q & 1 & 1* & H91\\
B3 0731+438 & 2.43 & G & 1 & 1 & H06\\
Q 0805+046 & 2.89 & Q & 1 & 1* & H91\\
TXS 0828+193 & 2.57 & G & 1 & 1 & H06\\
OTL 0852+192 & 2.47 & G & 0 & 0* & GK95\\
B2 0902+343 & 3.38 & G & - & - & R03\\
Q 0941+261 & 2.91 & Q & 1 & 1* & H91\\
TXS 0943-242 & 2.92 & G & 1 & 1 & H06\\
PKS 1138-262 & 2.12 & BG & 0 & 0 & P97; H07\\
6C 1232+39 & 3.22 & G & 1 & 1* & E93\\
MRC 1243+036 & 3.57 & G & 1 & 1* & vO96; H06\\
Q 1318+113 & 2.17 & Q & 1 & 1* & H91\\
Q 1345+584 & 2.04 & Q & 1 & 1* & H91\\
3C 294 & 1.79 & G & 1 & 1 & MC90\\
4C-00.54 & 2.36 & G & 1 & 1 & H06\\
8C 1435+635 & 4.25 & G & 1 & 1* & S95\\
TX 1436+157 & 2.55 & Q & 1 & 1* & vO97\\
3C 326.1 & 1.82 & G & 1 & 1* & MC87b\\
TXS 1558-003 & 2.52 & BG & 1 & 1 & H06\\
4C+10.48 & 2.35 & G & 0 & - & vO97; H07\\
4C+48.48 & 2.35 & G & 1 & 1 & C96; H06\\
MRC 2104-242 & 2.49 & G & - & - & P01; VM06\\
4C+23.56 & 2.49 & G & 1 & 1 & K97; H06\\
MG 2141+192 & 3.59 & G & 0 & - & M02\\
Q 2150+053 & 1.98 & Q & 1 & 1* & H91\\
Q 2222+051 & 2.32 & Q & 1 & 1* & H91\\
B3 J2330+3927 & 3.09 & G & 1 & - & PT05\\
Q 2338+042 & 2.59 & Q & 1 & 1* & H91\\
\hline
Total & & & 26/33 & 24/27 &\\
\hline
\end{tabular}
\end{table}

\renewcommand\thetable{\bf C\arabic{table}}
\setcounter{table}{0}

\section*{Appendix C: correlations with jet-sidedness in the parent sample}

In view of the possible correlation we found between jet-sidedness, polarization, brightness and spectral index of the radio source ($\S$3), we have investigated whether the wider sample of HzRG from which our sample was taken (Carilli et al. 1997; Pentericci et al. 2000b) shows a similar correlation with jet-sidedness.  We have considered all the sources from Carilli et al. (1997) and Pentericci et al. (2000b) that show apparent one-sided radio jets (see $\S3$ for our definition of a 'jet').  The results are presented in Table C1.  

We find similar results to those we reported in $\S$3: (i) the radio polarization is always highest on the jet-side (12/12 sources); (ii) the hotspot on the jet-side always has a flatter spectral index (13/13 sources); (iii) the hotspot on the jet-side is usually the brightest (12/15 sources); and (iv) there is no strong correlation between the arm-length asymmetry and the jet-sidedness, though there may be a weak trend for the longest lobe to be associated with the radio jet (9/14 sources).  

\begin{table}
\centering
\caption{Table illustrating the relationship between radio asymmetries for all sources from the parent HzRG sample of Carilli et al. (1997) and Pentericci et al. (2000b) that show a one-sided jet.  Columns: (1) Source name; (2) Source redshift z; (3) Parameter indicating whether the highest radio polarization is on the side of the apparent jet; (4) Parameter indicating whether the hotspot on the side of the apparent jet has the flattest radio spectrum (highest $\alpha$$_{R}$) between observed frequencies of 4.7 and 8.2 GHz; (5) Parameter indicating whether the brightest radio hotspot is on the side of the apparent jet; (6) Parameter indicating whether the shortest radio lobe is associated with the apparent jet.  A dash (-) denotes that there is no apparent asymmetry in a property.}
\begin{tabular}{llllll}
\hline
Source & z & J-Pol & J-$\alpha$$_{R}$ & J-R & J-Q\\  
(1) & (2) & (3) & (4) & (5) & (6)\\ \hline
MRC 0156-252 & 2.09 & - & 1 & 1 & 1\\
TXS 0211-122 & 2.34 & 1 & 1 & 1 & 1\\
MRC 0406-244 & 2.44 & 1 & 1 & 1 & 0\\
TXS 0417-181 & 2.77 & - & 1 & 1 & 0\\
B3 0731+438 & 2.43 & 1 & 1 & 1 & 0\\
TXS 0828+193 & 2.57 & 1 & - & 0 & 1\\
TXS 1113-178 & 2.34 & 1 & 1 & 1 & 1\\
PKS 1138-262 & 1.26 & - & 1 & 0 & 1\\
B3 1204+401 & 2.07 & 1 & 1 & 1 & 0\\
MRC 1243+036 & 3.57 & 1 & 1 & 1 & 0\\
4C 1345+245 & 2.89 & 1 & 1 & 1 & 0\\
4C-00.54 & 2.36 & 1 & 1 & 1 & 0\\
TXS 1558-003 & 2.52 & 1 & 1 & 1 & 0\\
6C 1908+722 & 3.54 & 1 & - & 0 & 0\\
MRC 2025-218 & 2.63 & 1 & 1 & 1 & -\\
\hline
Total & & 12/12 & 13/13 & 12/15 & 5/14\\
\hline
\end{tabular}
\end{table}

\end{document}